\documentclass[nohyper,notoc]{JHEP3}

\usepackage{amsmath,euscript,array,amssymb,cite,bm} 
\usepackage{graphicx}       
\def\cN{{\cal N}}

\def\geqsim{\stackrel{>}{\sim}}

\def\Tr{{\rm Tr}}
\def\sst{\scriptscriptstyle}
\def\det{{\rm det}}

\def\Dbarslash{\,\,{\raise.15ex\hbox{/}\mkern-12mu {\bar\D}}}
\def\Dslash{\,\,{\raise.15ex\hbox{/}\mkern-12mu \D}}
\def\delslash{\,\,{\raise.15ex\hbox{/}\mkern-9mu \partial}}
\def\delbarslash{\,\,{\raise.15ex\hbox{/}\mkern-9mu {\bar\partial}}}

\def\hf{{\textstyle{1\over2}}}

\def\D{{\cal D}}
\def\Dbarslash{\,\,{\raise.15ex\hbox{/}\mkern-12mu {\bar\D}}}
\def\delslash{\,\,{\raise.15ex\hbox{/}\mkern-9mu \partial}}
\def\Dslash{\,\,{\raise.15ex\hbox{/}\mkern-12mu \D}}

\def\={\, =\, }
\def\+{\, +\, }
\def\-{\, -\, }

\newcommand{\be}{\begin{equation}}
\newcommand{\ee}{\end{equation}}
\def\bea{\begin{eqnarray}}
\def\eea{\end{eqnarray}}

\title{SUSY breaking by a metastable ground state: Why the early Universe preferred the non-supersymmetric
vacuum}

\author{Steven A.~Abel, Chong-Sun Chu, Joerg Jaeckel and Valentin V.~Khoze\\
Centre for Particle Theory, University of Durham,
Durham, DH1 3LE, UK\\
{\tt s.a.abel@durham.ac.uk,}\, 
{\tt chong-sun.chu@durham.ac.uk,}\, 
{\tt jjaeckel@mail.desy.de,}\, 
{\tt valya.khoze@durham.ac.uk}}

\abstract{Supersymmetry breaking in a metastable vacuum is re-examined in a cosmological context. It is 
shown that thermal effects generically drive the Universe to the 
metastable minimum even if it begins in the supersymmetry-preserving one. This 
is a generic feature of the ISS models of metastable supersymmetry breaking 
due to the fact that SUSY preserving vacua contain 
fewer light degrees of freedom than the metastable ground state at the origin. 
These models of metastable SUSY breaking are thus placed on an 
equal footing with the more usual dynamical SUSY breaking scenarios.   }
 
\preprint{{\tt hep-th/0610334}
\\IPPP/06/76\\
DCPT/06/152
}

\begin{document}

\def\Nf{{N_{f}}}
\def\Nc{{N_{c}}}
\def\p{\varphi}
\def\pt{\tilde\varphi}
\def\uno{\mbox{1 \kern-.59em {\rm l}}} 
\def\vplus{|{\rm vac}\rangle_+}
\def\vzero{|{\rm vac}\rangle_0}

\section{Introduction}

It has recently been noted that
dynamical supersymmetry breaking in a long-lived metastable vacuum is
an interesting alternative to the usual dynamical supersymmetry
breaking scenario
\cite{ISS}\footnote{In a simple modified O'Raifeartaigh model the 
possibility of SUSY breaking in a metastable vacuum has been considered 
already a while ago in \cite{Ellis:1982vi}. 
It is interesting to note that one of the motivations for this work was that the Universe
may get stuck in such a metastable state.}. 
It was shown that a microscopic asymptotically free
theory can, for certain choices of parameters, 
be described by a Seiberg dual theory in the infra-red which has a metastable 
minimum at the origin. Moreover, the global supersymmetric 
minima are also located where the macroscopic theory is well under control, and it
could be shown that the tunnelling rate to the true global minima could
be made parametrically small.
This class of theories has recently attracted a lot of interest
\cite{F,O,B,A,F2,Forste,A2,Bena,Dine,Ahn,Eto,A3}.
It opens up new avenues for SUSY phenomenology,
but one open question is that of naturalness; can one
explain the fact that the Universe today resides in a metastable
minimum? Are these models more, or less, natural than the usual 
scenario of dynamical supersymmetry breaking?

In this paper we point out that ending up in such a
metastable minimum is in fact very natural and even generic in the context
of a thermal Universe. The theory at the metastable minimum 
has (quite generally)
more massless and nearly massless degrees of freedom than at the
supersymmetric minimum, an unusual consequence of the fact that the
SUSY preserving vacua appear non-perturbatively\footnote{In other words,
supersymmetry is {\em restored dynamically}.} in the Intriligator-Seiberg-Shih (ISS) model. 
We will show that due to these
relatively light states, at high temperatures (greater than the supersymmetry
breaking scale), the metastable minimum at the origin becomes the
global minimum. Moreover, we will see that even if the fields begin in the supersymmetric
minimum, high enough temperatures will thermally drive them to the 
SUSY breaking minimum where they remain trapped as the
Universe cools. 
More precisely, we consider a scenario where, at the end of inflation, the Universe is very cold
and we may assume that it is in the energetically preferred supersymmetric 
vacuum. 
Then the Universe reheats. If the temperature is high enough it will then automatically evolve towards the
SUSY breaking state as we will show in the following.
Thus these models offer an alternative and appealing explanation 
for {\em why} supersymmetry is broken: the early Universe was driven
to a supersymmetry breaking, metastable minimum by thermal effects.\footnote{If the 
initial conditions are such that the Universe
{\it starts} in the supersymmetry breaking state (the case considered in
Refs.~\cite{Craig:2006kx,Fischler:2006xh} subsequent to this paper), it will stay
there (cf. also \cite{ISS,Abel:2006my}).} 

In general we think of the settings where the ISS model forms a sector of the full theory
which includes a supersymmetric Standard Model (MSSM). Supersymmetry of the full theory is broken
if the ISS sector ends up in the metastable non-supersymmetric vacuum. For the full theory to be driven
to this SUSY breaking vacuum by thermal effects we, of course, have to assume that the relevant fields
of the supersymmetry breaking sector (ISS), of the MSSM sector, and of the messenger sector were at some time
in {\it thermal equilibrium}. This is the case if the SUSY breaking scenario is gauge mediation,
direct mediation, or even a visible sector breaking. (On the other hand, a SUSY breaking sector which couples to 
the MSSM
only gravitationally would remain out of thermal equilibrium.)

The fact that the Universe can be driven to a metastable minimum by
thermal effects is well known in the context of charge and colour
breaking minima in the MSSM \cite{original-idea,thermal}. However 
these models differ from our case in two respects. First, it is in the global minimum
away from the origin that a symmetry, namely supersymmetry, is being restored in our scenario. 
In addition the effective potential is extremely flat because the SUSY preserving true 
vacua are generated dynamically. It is therefore automatically 
very sensitive to finite temperature effects, whilst having a metastable 
vacuum that is extremely long lived. 

In this paper it will be sufficient to consider the ISS metastable SUSY breaking sector
on its own. The thermal transition to the non-supersymmetric vacuum will be driven by the above
mismatch between the numbers of light degrees of freedom in the two vacua of the ISS model.

\section{The ISS model at zero temperature} 

\subsection{Set-up of the model}

The Intriligator-Seiberg-Shih model \cite{ISS} is described by a supersymmetric
$SU(N)$ gauge theory coupled to $\Nf$ flavours of chiral superfields $\p^c_i$ and $\pt^i_c$ transforming in the
fundamental and the anti-fundamental representations of the gauge group; $c=1,\ldots,N$ and $i=1,\ldots, \Nf.$
There is also an
$\Nf \times \Nf$ chiral superfield $\Phi^i_j$ which is a gauge singlet.
The number of flavours is taken to be large, $\Nf  > 3N,$ such that the $\beta$-function for the gauge coupling is positive,
\be
b_0 = 3N - \Nf < 0
\label{bzero}
\ee
the theory is free in the IR and strongly coupled in the UV where it develops a Landau pole at the energy-scale $\Lambda_L$.
The Wilsonian gauge coupling is
\be
e^{-8 \pi^2 / g^2(E)} \, =\, \left(\frac{E}{\Lambda_L}\right)^{\Nf-3N}
\label{g2e}
\ee
The condition \eqref{bzero} ensures that the theory is weakly coupled at scales $E \ll \Lambda_L$, 
thus its low-energy dynamics as well as the vacuum structure is under control.
In particular, this guarantees a robust understanding of the theory
in the metastable SUSY breaking vacuum found in \cite{ISS}. This is one of the key features of the ISS model(s).

One notes that this formulation of the theory can only provide a low-energy effective field theory description
due to the lack of the asymptotic freedom in \eqref{bzero},\eqref{g2e}. 
At energy scales of order $\Lambda_L$ and above, this effective description breaks down and one
should use instead a different (microscopic) description of the theory, assuming it exists.
Fortunately -- and this is the second key feature of the ISS construction \cite{ISS} -- 
the ultraviolet completion of this effective theory is known and is provided by its Seiberg dual formulation
\cite{Seiberg1,Seiberg2,IS}. The microscopic description of the ISS model is
$\cN=1$ supersymmetric QCD with the gauge group 
$SU(\Nc)$ and $\Nf$ flavours of fundamental and anti-fundamental quarks $Q_i$ and $\tilde{Q}^i.$ 
The number of colours in the microscopic theory is $\Nc = \Nf -N$ and 
the number of flavours $\Nf$ is the same in both descriptions.
It is required to be in the range $\Nc +1 \le \Nf < \frac{3}{2} \Nc.$ 
The lower limit on $\Nf$ ensures that Eq.~\eqref{bzero} holds, while the $b_0^{\rm micro}$ coefficient
of the microscopic theory is positive (the $\beta$-function is negative and the microscopic theory is 
asymptotically free)
\be
 b_0^{\rm micro} \, = 3\Nc - \Nf \ , \qquad   
 \frac{3}{2} \Nc\, < \, b_0^{\rm micro}\, \le \, 2\Nc -1
\label{bzmicro}
\ee
This microscopic formulation of the ISS model will be referred to as the
$SU(\Nc)$, $\Nf$ {\it microscopic Seiberg dual}. It is weakly coupled in the UV and strongly coupled in the IR.
However, we already mentioned that
the vacuum structure of the theory and the supersymmetry breaking by a metastable vacuum 
should be considered in the low-energy effective description of the theory, which is IR free. 
This description  of the ISS model is known 
as the $SU(N)$, $\Nf$ {\it macroscopic Seiberg dual} formulation. From now
on we will always assume this macroscopic dual description of the ISS
model. For the purposes of this paper, the microscopic dual description is
only necessary to guarantee the existence of the theory above the Landau
pole $\Lambda_L.$

The tree-level superpotential of the ISS model is given by
\be
W_{\rm cl}\, =\, h\, \Tr_{\sst \Nf} \p \Phi \pt\, -\, h\mu^2\, \Tr_{\sst \Nf} \Phi
\label{Wcl}
\ee
where $h$ and $\mu$ are constants. The usual holomorphicity arguments imply that the superpotential
\eqref{Wcl} receives no corrections in perturbation theory. However, there is a non-perturbative contribution
to the full superpotential of the theory, $W=W_{\rm cl} + W_{\rm dyn},$ which 
is generated dynamically. $W_{\rm dyn}$ was determined in \cite{ISS} and is given by
\be
W_{\rm dyn}\, =\, N\left( h^\Nf \frac{\det_{\sst \Nf} \Phi}{\Lambda_{L}^{\Nf-3N}}\right)^\frac{1}{N}
\label{Wdyn}
\ee
This dynamical superpotential is exact, its form is uniquely determined by the symmetries of the theory
and it is generated by instanton-like configurations. 

The authors of \cite{ISS} have studied the vacuum structure of the theory and established the
existence of the metastable vacuum $\vplus$ characterised by
\be
\langle \p \rangle =\, \langle \pt^T \rangle = \, \mu \left(\begin{array}{c}
\uno_{N}\\ 0_{\Nf-N}\end{array}\right) \ , \quad
\langle \Phi \rangle = \, 0 \ , 
\qquad V_+ = \, (\Nf-N)|h^2 \mu^4| 
\label{vac+}
\ee 
where $V_+$ is the classical energy density in this vacuum. Supersymmetry is broken 
since $ V_+ >0.$ In this vacuum the $SU(N)$ gauge group is Higgsed 
by the vevs of $\p$ and $\pt$ and the gauge degrees of freedom
are massive with $m_{\rm gauge} = g \mu.$ 

This supersymmetry breaking vacuum $\vplus$ originates from the so-called
rank condition, which implies that there are no solutions to the F-flatness
equation $F_{\Phi^j_i}=0$ for the classical superpotential $W_{\rm cl}$ in \eqref{Wcl},
\be
F_{\Phi^j_i} =\, 
h(\pt^j_c \p_i^c - \mu^2 \delta^J_i) \neq \, 0
\label{rank}
\ee
The non-perturbative superpotential of \eqref{Wdyn} gives negligible contributions to the
effective potential around this vacuum and can be ignored there. It was further argued in \cite{ISS}
that the vacuum \eqref{vac+} has no tachyonic directions, is classically stable, and quantum-mechanically
is long-lived.

In addition to the metastable SUSY breaking vacuum $\vplus$, 
the authors of \cite{ISS} have also
identified the SUSY preserving stable vacuum\footnote{In fact there are precisely $\Nf-N=\Nc$ of such vacua
differing by the phase $e^{2\pi i/(\Nf-N)}$ as required by the Witten index of the microscopic Seiberg dual formulation.}
$\vzero$,
\be
\langle \p \rangle =\, \langle \pt^T \rangle = \, 0 \ , \quad
\langle \Phi \rangle = \, \Phi_0=\, \mu \gamma_0 \, \uno_{\Nf} \ , \qquad \qquad V_0 = \, 0
\label{vac0}
\ee
where $V_0$ is the energy density in this vacuum and
\be
\gamma_0 =\,\biggl( h \epsilon^{\frac{\Nf-3N}{\Nf-N}}\biggr)^{-1} \ , \quad {\rm and} \quad
\epsilon :=\, \frac{\mu}{\Lambda_L} \ll \, 1
\label{gamma0}
\ee
This vacuum was determined in \cite{ISS} by solving the F-flatness
conditions for the complete superpotential $W=W_{\rm cl}+W_{\rm dyn}$ of the theory.
In the vicinity of $\vzero$ the non-perturbative superpotential of \eqref{Wdyn}
is essential as it alleviates the rank condition \eqref{rank} and allows to solve
$F_{\Phi^j_i}=0$ equations. Thus, the appearance of the SUSY preserving vacuum
\eqref{vac0} can be interpreted in our macroscopic dual description as a non-perturbative
or dynamical restoration of supersymmetry \cite{ISS}.

When in the next Section we put the ISS theory at high temperature
it will be important to know the number of light and heavy degrees of freedom
of the theory as we interpolate between $\vplus$ and $\vzero$. The macroscopic description of the ISS
model is meaningful as long as we work at energy scales much smaller than the cut-off scale $\Lambda_L.$
The requirement that $\epsilon = \mu/\Lambda_L \ll 1$ is a condition that the gauge theory
is weakly coupled at the scale $\mu$ which is the natural scale of the macroscopic theory.
Equation \eqref{gamma0} implies that $\gamma_0 \gg 1$ or in other words that there is a natural
separation of scales $\mu \ll \Phi_0=\mu \gamma_0 \ll \Lambda_L$ dictated by  $\epsilon\ll 1.$
The masses of order $h\Phi_0$ will be treated as heavy
and all other masses suppressed with respect to $h\Phi_0$ by a positive power of $\epsilon$ we will refer to as light.
In the SUSY preserving vacuum $\vzero$ the heavy degrees of freedom are the $\Nf$ flavours of
$\p$ and $\pt,$ they get the tree-level masses $m_\p = h \Phi_0$ via \eqref{Wcl}.
All other degrees of freedom in $\vzero$ are light.\footnote{Gauge degrees of freedom 
in $\vzero$ are confined, but the appropriate mass gap, given by the gaugino condensate,
is parametrically (in $\epsilon$) smaller than $m_\p = h \Phi_0.$ Thus the gauge degrees of freedom
can still be counted as light for sufficiently high temperatures.}

In the vacuum $\vplus$ all of the original degrees of freedom of the ISS
theory are light. Hence we note for future reference that the SUSY
preserving vacuum $\vzero$ has fewer light degrees of freedom than the
SUSY breaking vacuum $\vplus$. The mismatch is given by the $N_f$
flavours of $\p$ and $\pt$.
The fact that the metastable SUSY breaking vacuum has fewer degrees of freedom than the 
supersymmetric vacuum
is an important feature of the ISS model. As we will
explain below, this property will give us a necessary condition for the thermal Universe to end up
in the metastable vacuum in the first place. 

\subsection{Effective potential}

Having presented the general set-up, let us now turn to the effective 
potential between the two vacua $\vplus$ and $\vzero$ of the ISS model. 
This can be determined in its entirety as follows.
We parameterise the path interpolating between the two vacua \eqref{vac+} 
and \eqref{vac0} in field space via 
\be
 \p(\sigma)  =\,  \pt^T (\sigma) = \,\sigma \mu\, \left(\begin{array}{c}
\uno_{N}\\ 0_{\Nf-N}\end{array}\right) \ , \quad
\Phi (\gamma) = \, \gamma \mu \uno_\Nf \ , 
\qquad  0 \le \gamma \le \gamma_0\ , \   1 \ge \sigma \ge 0
\label{inter}
\ee 
Since the K\"ahler potential in the free magnetic phase in the IR is that of the
classical theory, the effective potential $V$ can be determined from knowing the superpotential of the theory 
\eqref{Wcl}, \eqref{Wdyn}.
First we will calculate the classical potential $V_{\rm cl}$ along path \eqref{inter} and ignoring the non-perturbative
contribution $W_{\rm dyn}.$ Using \eqref{Wcl} we find,
\be
\frac{1}{|h|^2 } V_{\rm cl}(\gamma,\sigma) =\, \Tr_{\sst \Nf} |\Phi\p|^2 + 
\Tr_{\sst \Nf} |\pt \Phi|^2
+\Tr_{\sst \Nf} |\pt_i \p_j -\mu^2 \delta_{ij}|^2 =\,
|\mu|^4(2 N \gamma^2 \sigma^2 + N(\sigma^2-1)^2 +\Nf-N)
\label{Vcl1}
\ee
This expression for $V_{\rm cl}(\gamma,\sigma)$ is extremized for $\sigma=\sqrt{1-\gamma^2}$ or $\sigma=0.$
We substitute the first solution for $\sigma$ into $V_{\rm cl}$ for the range of $0\le \gamma \le 1,$ and the
second solution, $\sigma=0$ for $1\le \gamma \le \gamma_0.$ 
We thus obtain
\begin{eqnarray}
\frac{1}{|h^2 \mu^4| }\,  V_{\rm cl}(\gamma) & = & \left\{ \begin{array}{cc}
\Nf-N + \,2 N \gamma^2(1- \hf\gamma^2)  & \qquad 0 \le \gamma\le 1 \\
\Nf  & \qquad 1\leq\gamma\leq\gamma_{0}\\
\end{array}\right.\end{eqnarray}
This potential is accurate for $\gamma\le 1$, where the potential is rising. In order to address the flat regime
$1\leq\gamma\leq\gamma_{0}$ we need to include the contribution of the non-perturbative $W_{\rm dyn}.$
This gives 
\be
V =\, \Tr_{\sst \Nf} \left|\frac{\partial}{\partial \Phi_{ij} }
N \left( h^\Nf \frac{\det_{\sst \Nf} \Phi}{\Lambda_{L}^{\Nf-3N}}\right)^\frac{1}{N}
-h \mu^2 \delta_{ij}\right|^2 =\, 
|h^2 \mu^4|\, \Nf\biggl( \bigl(\frac{\gamma}{\gamma_0}\bigr)^{\frac{\Nf-N}{N}}-1\biggr)^2
\ee
Combining these expressions we have the final result for the effective potential
interpolating between the two vacua:
\begin{eqnarray}
 \hat{ V}_{T=0} (\gamma) \equiv \, \frac{1}{|h^2 \mu^4| }\,  V(\gamma)_{T=0}\, 
& = & \left\{ \begin{array}{cc}
\Nf-N + \,2 N \gamma^2(1- \hf\gamma^2)  & \qquad 0 \le \gamma\le 1 \\
\\
 \Nf\biggl( \bigl(\frac{\gamma}{\gamma_0}\bigr)^{\frac{\Nf-N}{N}}-1\biggr)^2
& \qquad 1\leq\gamma \\
\end{array}\right.
\label{VzeroT}
\end{eqnarray}
For our forthcoming thermal applications we have included
the subscript $T=0$ to indicate that this effective potential is calculated at zero temperature.
We plot the zero-temperature effective potential \eqref{VzeroT} in Figure 1.
\begin{figure}[t]
    \begin{center}
        \includegraphics[width=8cm]{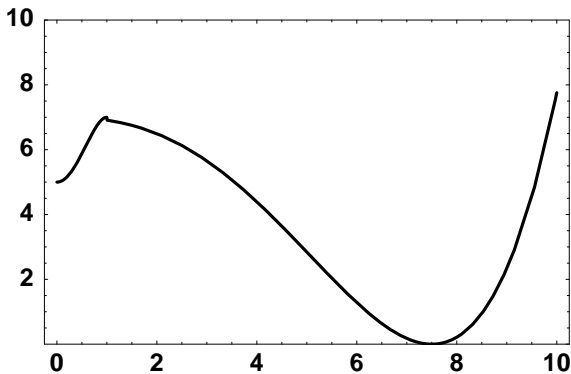}
    \end{center}
    \caption{Zero temperature effective potential $\hat{ V}_{T=0} (\gamma)$ of Eq.~\eqref{VzeroT} as a function of 
    $\gamma = \Phi/\mu.$ For the SUSY preserving vacuum $\vzero$ we chose $\gamma=\gamma_0=7.5$.
    The SUSY breaking metastable minimum $\vplus$ is always at $\gamma=0,$ and the top of the barrier is 
    always at $\gamma=1.$  We have taken the minimal allowed values for $N$ and $\Nf$, $N=2$, $\Nf=7.$ }
    \label{fig:recurs}
\end{figure}
The key features of this effective potential are (1) the large distance between the two vacua,
$\gamma_0 \gg 1,$ 
and (2) the slow rise of the potential to the left of the SUSY preserving vacuum.
(For esthetic reasons $\gamma_0$ in Figure 1. is actually chosen to be rather small, $\gamma=\gamma_0=7.5.$)

We note that
the matching between the two regimes at $\gamma =1$  in \eqref{VzeroT} holds
up to an insignificant $1/\gamma_0^{\Nf-N \over N}$ correction which is small since
$\gamma_0 \gg 1.$ This tiny mismatch (which can be seen in Figure 1.)
can be easily corrected in the derivation of $V$ by using the full superpotential
in both regimes, rather than neglecting $W_{\rm dyn}$ at $\gamma \le 1.$ However, it will not change 
anything in our considerations. Other corrections which we have dropped from the final expression include
perturbative corrections to the Kahler potential. These effects are also expected to be small at scales much below
the cut-off $\Lambda_L.$

The authors of \cite{ISS} have already estimated the tunnelling rate from the metastable $\vplus$
to the supersymmetric vacuum $\vzero$ by approximating the potential in Figure 1 in terms of a triangle.
The action of the bounce solution in the triangular potential is of the form,
\be
S^{4D}_{\rm bounce} =\, \frac{2 \pi^2}{3 h^2}\frac{N^3}{\Nf^2} \, \left(\frac{\Phi_0}{\mu}\right)^4 \sim \, 
\frac{h^{-6}}{\epsilon^{4(\Nf-3N)/(\Nf-N)}}
\, \gg 1
\qquad {\rm for} \quad \epsilon \ll 1
\ee
Thus as argued in \cite{ISS} it is always possible, by choosing sufficiently 
small $\epsilon$, to ensure that the decay time of the metastable vacuum 
to the SUSY ground state is much longer than the age of the Universe. 

On closer inspection the constraints imposed by this condition are 
in any case very weak. In order to estimate the required value of $\epsilon$,
note that the Euclidean action $S^{4D}_{\rm bounce}$ gives
the false vacuum decay rate per unit volume as
\begin{equation}
\Gamma_{4}/V=D_{4}e^{-S^{4D}_{\rm bounce}}.
\end{equation}
$D_{4}$ is the determinant coefficient and is irrelevant to the discussion.
To determine whether the Universe could have decayed (at least once)
in its lifetime, we multiply the above by the space-time volume of
the past light-cone of the observable Universe. This results in a
bound to have no decay of roughly \cite{thermal,nonthermal} 
\be
S^{4D}_{\rm bounce} \geqsim 400.
\ee
This translates
into an extremely weak lower bound on $\Phi_{0}$,
\be
\left(\frac{\Phi_{0}}{\mu}\right)\, \geqsim \, 3 \sqrt{h}\,  
\left(\frac{\Nf^2}{N^3}\right)^{\frac{1}{4}}.
\label{vacstab}
\ee
Note that the bound is on the relative width to height of the bump. Thus it is 
indeed simply the flatness of the potential which protects the 
metastable vacuum against decay. 
In terms of $\epsilon=\mu/\Lambda_L$ the bound \eqref{vacstab}
reads
\be
\epsilon^{1-\frac{2N}{\Nf-N}} \, \lesssim\, 
\frac{1}{3 h\sqrt{h} }\, \left(\frac{N^3}{\Nf^2}\right)^{\frac{1}{4}}.
\label{vacstab2}
\ee
It is interesting to note that this very weak bound \eqref{vacstab}, \eqref{vacstab2}
becomes strong when expressed in terms of $\epsilon$ itself in the minimal case
of $N=2$ and $\Nf=7$,
\be
\epsilon \, \lesssim\, 
\left(\frac{1}{3 h\sqrt{h} }\, \left(\frac{8}{49}\right)^{\frac{1}{4}}\right)^5 \,
\simeq \, 4.3 \cdot 10^{-4}\, \left(\frac{1}{h}\right)^{\frac{15}{2}} \ll 1.
\label{vacstab3}
\ee
In the following Section we will explain why at high temperatures 
the Universe has ended up in the metastable non-supersymmetric vacuum 
in the first place.

\section{Dynamical evolution at finite temperature} 

\subsection{The shape of the effective potential at finite temperature}

The effective potential at finite
temperature along the $\Phi$ direction is governed by the following well-known expression \cite{Vthermal}:
\be
V_{T}(\Phi)=\, V_{T=0}(\Phi)+\frac{T^{4}}{2\pi^{2}}\sum_{i}\pm n_{i}\int_{0}^{\infty}\mbox{d}q\, 
q^{2}\ln\left(1\mp\exp(-\sqrt{q^{2}+m_{i}^{2}(\Phi)/T^{2}})\right)
\label{Vth1}
\ee
The first term on the right hand side is the zero temperature value of the effective potential;
for the ISS model we have determined it in Eq.~\eqref{VzeroT} and in Figure 1.
The second term is the purely thermal correction (which vanishes at $T=0$) and it is determined 
at one-loop in perturbation theory. The $n_i$ denote the numbers of degrees of freedom present in the 
theory\footnote{Weyl fermions and complex scalars each count as $n=2.$} 
and the summation is over all of these degrees of freedom. The
upper sign is for bosons and the lower one for fermions. Finally, $m_{i}(\Phi)$ denote the masses of these degrees
of freedom induced by the vevs of the field $\Phi$.

We have noted in the previous Section that as we interpolate from the metastable vacuum to the supersymmetric one,
the $\Nf$ flavours of $\p$ and $\pt$ acquire masses $m_\p= h\Phi =h\mu \gamma$
and become heavy at large values of $\Phi.$ To a good approximation all other degrees of freedom can be counted
as essentially massless.\footnote{In the companion paper \cite{Abel:2006my}
we will refine this analysis by including contributions of gauge degrees of freedom to \eqref{Vth1}.
This will lead to an even lower bound for the reheat temperature, $T_R$, necessary for the
Universe to end up in the non-supersymmetric $\vplus$.}
As we are not interested in the overall additive ($T$ dependent, but field independent) 
constant in the thermal potential,
we need to include in \eqref{Vth1} only those variables whose masses vary from being light in one vacuum to being heavy 
in the other. These are precisely the $\p$'s and $\pt$'s. This implies\footnote{
There are only two terms in the sum in \eqref{Vth2}: one for bosons (+), and one for fermions (-). 
The prefactor $4N\Nf$ counts the total number of bosonic degrees of freedom in $\p^a_i$ and $\pt^i_a.$}
\be
\hat{V}_{\Theta}(\gamma)=\,
\hat{V}_{\Theta=0}(\gamma)+\frac{h^{2}}{2\pi^{2}}\Theta^{4}\sum_{\pm}\pm 4N \Nf \int_{0}^{\infty}\mbox{d}q\, 
q^{2}\ln\left(1\mp\exp(-\sqrt{q^{2}+\gamma^{2}/\Theta^{2}})\right).
\label{Vth2}
\ee
The above expression is written in terms of a dimensionless variable $\gamma = \Phi/\mu$ and we have also defined
a rescaled temperature 
\be
\Theta=T/|h\mu|
\ee 
In Figure 2 we plot the thermal potential \eqref{Vth2} for a few characteristic values of temperature.
\begin{figure}[t]
    \begin{center}
        \includegraphics[width=8cm]{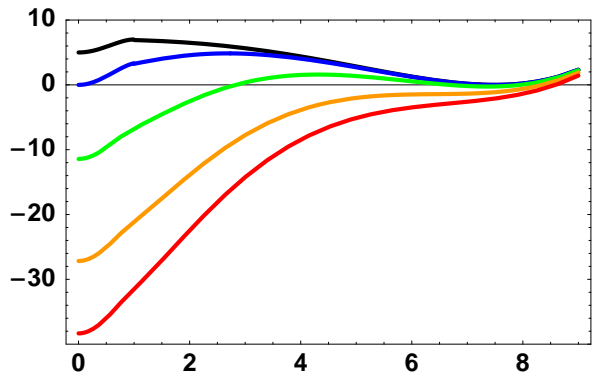}
    \end{center}
    \caption{Thermal effective potential \eqref{Vth2} for different values of the temperature. 
    Going from bottom to top, the red line corresponds to the temperature $\Theta \gtrsim \Theta_{\rm crit}$
    where we have only one vacuum at $\gamma=0$. The orange line 
    corresponds to $\Theta\approx \Theta_{\rm crit}$ where the second vacuum appears and the classical rolling stops. The green line is in the interval 
    $\Theta_{\rm degen} < \Theta < \Theta_{\rm crit}$
    where one could hope to tunnel under the barrier. The blue line is at  $\Theta \sim \Theta_{\rm degen}$
    where the two vacua become degenerate. Finally, the black line gives the zero temperature potential where
    the non-supersymmetric vacuum at the origin becomes metastable.}
    \label{fig:recurs2}
\end{figure}

In general one can consider a very wide range of temperatures $0\le T
\ll \Lambda_L $ corresponding to 
$0\le \Theta \ll 1/(h\epsilon)$. 
{}From Fig.
\ref{fig:recurs2} where we have plotted the effective potential at
various temperatures we immediately make out several interesting
temperatures. At high enough temperature there is only one minimum at
the origin and one would expect to be able to roll down classically to
the non-supersymmetric minimum $\vplus$,
as will be discussed in Section 3.2.3.
Below $\Theta=\Theta_{\rm crit}$ a second minimum forms. It will turn out, that
for our ISS potential this critical temperature will actually be less
than $\gamma_0$. 
In the following analysis, we will consider the range of
temperatures $h\mu<T \le h\Phi_{0}$ corresponding to $1<\Theta \le
\gamma_{0}$.

At temperatures below $\Theta_{\rm crit}$ the SUSY breaking minimum at the origin still has lower 
free energy then $\vzero.$
We will calculate the bubble nucleation rate to the true vacuum $\vplus$ in Section 3.2.2. 
Another distinguished temperature 
$\Theta_{\rm degen}$ is when the SUSY preserving vacuum $\vzero$
becomes degenerate with the non-supersymmetric vacuum $\vplus$. 
For a generic potential one would expect that somewhere
between these two scales 
there is a temperature $\Theta_*$
($\Theta_{\rm degen} < \Theta_* \le \Theta_{\rm crit}$)
where the rate of bubble nucleation turns from large to small. 
Our main goal is to determine
the temperature where the transitions (classical or tunnelling ones) from $\vzero$ to $\vplus$
stop.

\subsection{Transition to a SUSY breaking vacuum and the lower bound on $\bm T_R$}

In this subsection we discuss the two possibilities for the field to evolve to the SUSY breaking vacuum $\vplus$, 
classical evolution and bubble nucleation.

\subsubsection{Classical evolution to the SUSY breaking vacuum: Estimate of $T_{\rm crit}$}

It is clear from our expression for the thermal potential that at sufficiently high temperatures
there is only one minimum, namely $\vplus$, and the SUSY preserving minimum disappears.
This is caused by $\Nf$ flavours of $\p$ and $\pt$ becoming heavy away from the origin.
The disappearance of $\vzero$
is most easily seen analytically in the limit of very high temperatures
$\gamma \ll \Theta$ where the thermal potential grows as $\Theta^2 \gamma^2$ in the $\gamma$ i.e. $\Phi$ direction,
\be
\hat{V}_{\Theta} -
\hat{V}_{\Theta=0}(\gamma) \sim \, N \Nf \, \Theta^2 \gamma^2 - \,{\rm const}\,  \Theta^4 \ , \qquad
\Theta \gg \gamma = \Phi /\mu
\ee
Thus there is a critical temperature $\Theta_{\rm crit}$ such that at $\Theta > \Theta_{\rm crit}$ there is
only one minimum, and at $\Theta < \Theta_{\rm crit}$ the supersymmetry preserving second minimum starts to materialise.

A better estimate can be obtained by using the approximation,
\begin{equation}
\label{approx}
\pm \int_{0}^{\infty}\mbox{d}q\, 
q^{2}\ln\left(1\mp\exp(-\sqrt{q^{2}+\gamma^{2}/\Theta^{2}})\right)
\sim 
\Theta^4 \left(\frac{\gamma}{ 2 \pi \Theta}\right)^{\frac{3}{2}}
\exp\left(- \frac{\gamma}{\Theta}\right),\quad \rm{for}\quad \gamma\gg \Theta 
\end{equation}
for the integral in the $\Theta$ dependent part of Eq. \eqref{Vth2}. 
Comparing the derivatives (first and second)
of the $\Theta$-dependent part and the $\Theta$-independent part of the effective potential $\hat{V}_{\Theta}$
 in the vicinity of $\gamma_{0}$ we  get an estimate,
\begin{equation}
 \Theta_{\rm crit}\approx\,  \frac{\gamma_{0}}{\log\left({\gamma^{4}_{0}}/{C}\right)}\ ,  \qquad 
 C=\, \frac{\sqrt{2}\pi^{\frac{3}{2}}}{N h^2}\left(1-\frac{\Nf}{N}\right)^{2}
\end{equation}
which we have also confirmed numerically.

Although the height of the barrier in the zero temperature potential is only of the order
of $\mu$ where $\mu\ll \gamma_{0}\mu=h\Phi_{0}$ the temperature necessary to erode
the second minimum is only slightly (logarithmically) smaller than
$h\Phi_{0}$ due to the exponential suppression 
$\sim \exp(-\gamma/\Theta)$ apparent in Eqs. \eqref{Vth2} and \eqref{approx}.

In the early Universe the situation is not static and the temperature decreases due to the expansion
of the Universe.
So one should check that even in the absence of a second minimum the field has time enough to evolve to $\vplus$ 
before the temperature drops below $\Theta_{\rm crit}$. We will comment on this issue more in the next subsection. 
Here, we just point out that the temperature drops on a time scale $M_{Pl}/T^2\gg 1/T$.

\subsubsection{Bubble Nucleation: Estimate of $T_*$}

Let us now turn to the possibility of bubble nucleation, with an estimate 
of the temperature $T_*$ where the transitions 
from $\vzero$ to $\vplus$ change from fast to slow. 
Bubble nucleation is of course only possible above the temperature where the 
two minima become degenerate, so let us first estimate this. 
We compare the effective potential (\ref{Vth2}) at the origin 
\be
\label{Vor}
\hat{V}_\Theta(\gamma=0) =\,  
{(N-N_f)}-
\frac{\pi^2 h^2 \Theta^4}{90}\left(n_B+\frac{7}{8}n_F\right) 
\ee
to the value of the effective potential at the second vacuum. The latter remains approximately
zero since the temperature $T_{\rm degen}$ is much below $\Phi_0.$
In \eqref{Vor}
 $n_B(n_F)$ counts the number of bosonic(fermionic) degrees of freedom ($4 N N_F$ in both cases). 
We thus find  
\begin{equation} \label{degen}
\Theta_{\rm degen}= \left({\frac{12 (N_f-N)}{h^2 \pi^{2} N N_f}}\right)^{\frac{1}{4}} ,\qquad \rm{for} 
\qquad 
\gamma_{0}\gg 1.
\end{equation}
Now we would like to know when bubble nucleation is fast enough
to lead to a phase transition. Thus we consider
temperatures in the range $\Theta_{\rm degen}<\Theta<\Theta_{\rm crit}$ such that the second vacuum $\vzero$ is already formed but still higher than $\vplus$. 
In the following we want to estimate $\Theta_*$ where tunnelling becomes too slow.

For this purpose we will now derive a simple estimate on the action of the tunnelling trajectory in the thermal
effective potential of the our model where the temperature is in the range  $\Theta_{\rm degen} < \Theta \le \Theta_{\rm crit}.$
The potential we want to model is depicted in Figure 2, its characteristic features are that the `false' vacuum
$\vzero$ is far away in the field space from $\vplus$ and that
the potential climbs very slowly from  $\vzero$ to a shallow barrier and then descends steeply to the `true' vacuum $\vplus$.
These features are a reflection of the fact that at zero temperature the vacuum $\vzero$ was generated non-perturbatively
(as reviewed in the previous Section). 
We will model this class of potentials with a simple linear no-barrier potential\footnote{We have checked the dependence on the characteristic scales of the potential using other simple approximations, for example two matched quadratics and a triangular potential.},
\be
\hat{V}_{\rm lin}(\gamma) =\,   \theta (\eta-\gamma)\, \Nf K \,(\gamma -\eta)
\label{Vsimple}
\ee
where $K$ and $\eta$ are constants, $\Nf$ is put for convenience, and $\theta$ is the step-function. We plot this potential
in Figure 2 for a convenient choice of constants along with the exact thermal potential.
It is clear from this figure that the tunnelling rate for our model potential will always be a little
higher than in the real case.

\begin{figure}[t]
    \begin{center}
        \includegraphics[width=8cm]{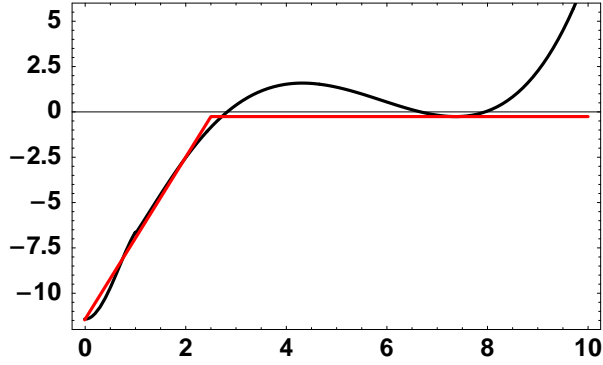}
    \end{center}
    \caption{A simple no-barrier model for the thermal effective potential. 
    The modelling potential $\hat{V}_{\rm lin}(\gamma)$
    is shown in red and the exact thermal potential $\hat{V}_{T}(\gamma)$is black.
    The tunnelling rate to the
    true vacuum for this model is calculated in the text.}
\end{figure}

The potential \eqref{Vsimple} admits a 
simple analytic solution for the bounce configuration. The tunnelling in this potential was discussed in detail in
Ref.~\cite{LW} in the $T=0$ case. We have performed a similar calculation in the 3-dimensional settings relevant to the
thermal case. The tunnelling configuration is the bounce solution $\gamma_{\rm bounce}(x)$ which extremizes
the 3-dimensional action
\be
 \frac{1}{T}\, 
S^{3D} =\,  \frac{1}{T}\, 4\pi
\int dr\, r^2 \left(\hf \Tr_{\Nf}(\Phi')^2 + V_T(\Phi)\right)
=\,
\Nf\, \frac{4\pi }{\Theta h^2}\, 
\int dx\, x^2 \left(\hf (\gamma')^2 + \frac{1}{\Nf}\, \hat{V}_{\rm lin}(\gamma)\right)
\label{Seq3d}
\ee
with the appropriate initial conditions.
The right hand side of \eqref{Seq3d} is written in terms of 
our usual dimensionless variables $\gamma,$ $\Theta$ and
the rescaled radial distance $x= h\mu r$. 
The classical equation for the bounce is then
\be
\gamma'' +\frac{2}{x} \gamma' = \, {\frac{1}{\Nf}}\, \partial_{\gamma} \hat{V}(\gamma)
\label{beq3d}
\ee
For the case \eqref{Vsimple} the bounce solution reads
\begin{eqnarray}
 \gamma_{\rm bounce}(x) 
& = & \left\{ \begin{array}{cc}
\frac{1}{6} Kx^2 
  & \qquad x \le x_m \\
 3\eta -  2\eta \frac{ x_m}{x}
& \qquad x \ge x_m  
\end{array}\right.
\label{bounceT}
\end{eqnarray}
where $x_m = \sqrt{\frac{6\eta}{K}}$ is the matching point for the two branches 
of the bounce solution and of its derivative.
The field configuration \eqref{bounceT} describes a bubble of size $x_m$ 
with $\gamma=3\eta$ on the outside
at large $x \gg x_m,$ and $\gamma=0$ on the inside at $x=0.$
The asymptotic value of $\gamma_{\rm bounce}$ is the `false' vacuum $\vzero$, thus
we should identify $3 \eta$ with $\gamma_0.$

The action \eqref{Seq3d} on the bounce trajectory  is
\be
\frac{1}{T}\, S^{3D}_{\rm bounce} =\, 
\frac{ 4\pi }{\Theta h^2} \, \Nf\,\frac{8 \sqrt{6}}{5} \, \frac{\eta^{2+\frac{1}{2}}}{K^{\frac{1}{2}}}
\ee
Now using the identification $3\eta= \gamma_0$ and $\Nf K\eta \propto \Delta \hat{V},$ where $\Delta \hat{V}$ is the 
drop in the potential which should be taken to be $\Delta \hat{V} \propto \Theta^4$ we get 
\be
\frac{1}{T}\, S^{3D}_{\rm bounce} \,\approx  {\rm const}\, \left(\frac{\Phi_0}{T}\right)^3
\label{bnest}
\ee
using the original dimensional variables $\Phi_0$ and $T$.
In order for the Universe to have undergone a phase transition by 
bubble nucleation one requires a sufficiently
high nucleation probability of bubbles of metastable vacuum. The nucleation
rate is given by \be
\Gamma_{3}\sim T^{4}e^{-S^{3D}/T}.\ee
This rate must be integrated from a maximum (reheat) temperature $T_{R}$
to the temperature $T_{\rm degen}\sim\mu$ at which the metastable and supersymmetric
minima are degenerate. The fraction of space remaining in the broken
phase is $e^{-P}$ where \cite{thermal}
\begin{eqnarray}
P & \sim & \frac{M_{Pl}^{4}}{T_{0}^{3}}\int_{T_{\rm degen}}^{T_{R}}\, T^{-2}(1-T_{0}/T)^{3}e^{-S^{3D}/T}dT\\
 & \approx & \frac{M_{Pl}^{4}}{3T_{0}^{3}S^{3D}}\, e^{-\frac{S^{3D}}{T_{R}}}\end{eqnarray}
where $T_0$ is the temperature of the Universe today. This gives 
\be
P\sim e^{256-\frac{S^{3D}}{T_{R}}}\,\frac{M_{W}}{S^{3D}}
\ee
and the bound becomes $\frac{\Phi_{0}}{T_R} \lesssim 1.3 $.
{}From this we conclude that the temperature $T_*$ at which the tunnelling 
transitions are still possible is very high (at the very top of its defining range),
\be
T_* \, \gtrsim \,  \Phi_0 \ , \qquad {\rm or} \qquad \Theta_* \, \gtrsim \gamma_0 
\ee
The temperature where bubble nucleation becomes significant has the same 
parametric dependence as the critical temperature $\Theta_{\rm crit}$ 
(our simple estimate is insufficient to capture logarithmic dependencies) 
and we conclude that $T_{*}\sim T_{\rm crit}$.

\subsubsection{Behaviour of the field after nucleation/rolling}

If the temperature rises above the critical temperature, the field
is in principle free to roll; however, is the critical temperature
a sufficient condition for the field to always end up at the origin?
Assuming that the Universe evolves in the standard FRW manner,
we should check that the phase transition has time to complete, and
that the falling temperature does not {}``overtake'' the field.
In order to model the field after the transition, we can approximate
the potential at temperatures above $T_{\rm crit}$ as 
linear\footnote{Here, we want just a rough estimate of the time scale
of the evolution towards the supersymmetry breaking minimum. For this we use a simple
adiabatic approximation (i.e. we use classical evolution equations but with an effective thermal potential) and a simplified potential. One can argue that, in general, some
kind of non-adiabatic treatment is required. Nevertheless, we expect that our conclusions from
this simple estimate remain qualitatively correct.}; 
\be
V =\, {\rm const}\,  T^4 \, \frac{\Phi}{\Phi_{0}} - {\rm const'}
\ee
Neglecting for the moment the effect of the $\varphi\Phi\tilde{\varphi}$
coupling, the field equations are \be
\ddot{\Phi}+3H\dot{\Phi} =\, - {\rm const}\,\frac{T^{4}}{\Phi_{0}}.
\ee
We may assume that the Hubble constant $H\sim T^{2}/M_{Pl}$ is negligible in this equation, 
so, taking initial
values $T_0 \gtrsim T_{\rm crit}$ i.e.
$T_{0}\sim\Phi_{0}$, we find that the field falls to the origin
in time 
\be
\Delta t_{\rm roll} \sim \,\frac{\Phi_{0}}{T_{0}^{2}} \sim \, \frac{1}{T_{0}}
\sim\,  M_{Pl}^{-\frac{1}{2}}t_{0}^{\frac{1}{2}}
\label{deltime}
\ee
where on the right hand side $t_0$ defines the time at which the temperature is $T_0,$
and we have considered a radiation dominated Universe with scale factor 
$a(t) \sim (t/t_0)^\hf.$
The result \eqref{deltime} should be compared to the typical time necessary for the Universe
to cool to temperatures where the second minimum $\vzero$ becomes prominent.
The evolution of $T$ is given by
\be
T(t) =\, \frac{T_0}{a(t)} \sim \, T_0 \, \left(\frac{t_0}{t}\right)^\hf
\ee
For example the time needed to reach $T = T_0/\sqrt{2}$  is 
$\Delta t_{\rm cool} = t_{0}$ which gives $\Delta t_{\rm cool} \gg\Delta t_{\rm roll}.$ This means that
the field is able to roll essentially instantaneously to the origin
where it undergoes coherent oscillation. 

Having concluded that the fields roll down to $\vplus$ sufficiently fast, we still need to
check that the oscillations of the fields around $\vplus$ would not bring us back to $\vzero$
as the Universe cools.
It is easy to see that if there was only the Hubble damping
the $\Phi$ field would oscillate out of the
origin as the Universe cools. Indeed, the $\Phi$ oscillations essentially
preserve energy in a comoving volume; \be
m_{\Phi}^{2}\Phi_{\rm max}^{2}R^{3}=\mbox{const~,}\ee
where $R\sim t^{\frac{1}{2}}$ is the scale factor in a radiation
dominated Universe, and $m_{\Phi}=\frac{hT}{2}$ is the temperature
induced mass of $\Phi$ at the origin. In an adiabatic regime $RT=const$
so we find \begin{eqnarray}
\frac{\Phi_{\rm max}(T)}{\Phi_{0}} & = & \sqrt{\frac{T}{T_{0}}}.\end{eqnarray}
The field would then escape from the origin because when $T\sim\mu$,
(assuming $T_{0}\sim\Phi_{0}$) the size of the oscillations would
be \be
\Phi_{\rm max}\sim\sqrt{\Phi_{0}\mu}\gg\mu.\ee
Fortunately, any open decay channel of $\Phi$ will typically provide a damping rate
$\Gamma_{\Phi}\sim T$, capturing the field in $\vplus$.
An example are couplings to the messenger sector fields, $f$, of the form $h^{\prime}\tilde{f}\Phi f.$
Since we assume thermal equilibrium, such a coupling must exist and cannot be arbitrarily weak.
The decaying oscillation
amplitudes are of order\begin{eqnarray}
\frac{\Phi_{\rm max}(T)}{\Phi_{0}} & = & \sqrt{\frac{T}{T_{0}}}e^{-\frac{1}{2}\Gamma_{\Phi}(t-t_{0})}.\end{eqnarray}
Sufficient damping always occurs; imposing
$\Phi_{\rm max}<\mu$ requires only that $\frac{M_{Pl}}{\Phi_{0}}\stackrel{>}{\sim}\log(\Phi_{0}/\mu)$
which is always satisfied
and the Universe ends up in the SUSY breaking state $\vplus.$ 
Note, however, that the number of oscillations
before damping is proportional to $\log(\Phi_{0}/\mu$) and may be
large.

\subsubsection{Lower bound on reheating temperature $T_R$}

Now, let us summarize our analysis 
and explain under which conditions it is natural for the early Universe to settle down at 
the metastable SUSY breaking vacuum. 
At the end of inflation the Universe is in a state of very low temperature and one may assume that it is in 
the energetically preferred supersymmetric state\footnote{If it is already in the supersymmetry breaking vacuum, $\vplus$, 
 it will stay there.} $\vzero$.
After inflation the Universe reheats to a temperature $T_{R}$. 
Already at relatively low temperatures $\sim T_{\rm degen}\sim\mu$ the supersymmetry breaking vacuum will 
have lower free energy than the supersymmetric vacuum.
However, if $T_R$ falls in the range $T_{\rm degen} <T_R\lesssim T_*\sim T_{\rm crit}$,  the Universe will remain in the state $\vzero$ 
although $\vplus$ is energetically preferred. It is stuck there because the barrier makes
classical evolution impossible and bubble nucleation is too slow.
Above $T_*$ the bubble nucleation rate will be high and above $T_{\rm crit}$ the classical evolution becomes possible. 
Hence, the Universe will evolve to the preferred supersymmetry breaking state $\vplus$.
We conclude that if the reheating temperature $T_R$ fulfills 
\be \label{cond-TR}
T_{\rm crit} \sim \frac{\mu}{ \epsilon^{1-2N/(N_f-N)} } \, \lesssim\,  T_R \, \lesssim  \,
\frac{\mu}{\epsilon}
\ee
it is ensured that the Universe 
always ends up in the supersymmetry breaking ground state\footnote{
In this estimate we ignore weak logarithmic corrections in $\epsilon$.}
$\vplus.$ 

In principle one should also consider the possibility of a transition
back towards the SUSY vacuum for $T< T_{\rm{degen}}$. It is known that this does not happen
at zero temperature \cite{ISS} . We have made a simple estimate of this effect at $T>0$ and concluded
that the Universe remains trapped. More recently, this has been discussed in depth in 
\cite{Craig:2006kx,Fischler:2006xh,Abel:2006my}.

The condition \eqref{vacstab} or \eqref{vacstab2} that the metastable vacuum $\vplus$
 is long-lived at zero temperature does not necessarily require $\Phi_0$
 to be very large or
 $\epsilon^{1-2N/(N_f-N)}$ to be very small.
For a low supersymmetry breaking
 scale $\mu$  of order of a few TeV
 the lower bound on the reheating temperature in \eqref{cond-TR} can be easily satisfied
for reheating temperatures as low as around 10 or 100 TeV.
On the other hand, for models with a significantly higher SUSY breaking scale,
this bound becomes more constraining. 

\section{Conclusions and discussion}

We have examined the consequences of meta-stable SUSY breaking 
vacua in a cosmological setting. As noted in \cite{ISS}, 
the ISS theory naturally admits a finite number of isolated 
supersymmetric vacua (as determined by the Witten index) along 
with a larger moduli space of  metastable SUSY breaking vacua. 
Since the latter is a much bigger configuration space, ISS suggested that 
it is more favorable for the Universe to be populated in the
metastable SUSY breaking vacua. In this paper, we have
considered thermal effects in the ISS model and 
have shown that the early Universe is always driven to metastability: 
as long as the SUSY breaking sector is in thermal equilibrium this 
provides a generic, dynamical explanation why supersymmetry became broken. 
This phenomenon is a consequence of two distinguishing properties of the 
ISS 
theory which are not necessarily shared by other models with 
metastable SUSY breaking vacua, namely, that the metastable minima
have more light degrees of freedom than the SUSY preserving vacua, 
and that the metastable vacua are separated from the SUSY preserving 
vacuum by a very shallow dynamically induced potential. 
Both play significant roles in our argument and our various estimates.

Given the current interest in "landscapes" of one variety or another, 
it is difficult to resist speculating on their implications for 
the ISS models, given the conclusions of our study. 
One longstanding problem that can be addressed in this context is that of 
the hierarchy between the Planck and supersymmetry breaking scales. 
Consider the possibility that the SUSY breaking sector has in 
fact a large number of product groups, with a landscape of metastable 
ISS minima into which the Universe could be thermally driven, with a 
range of values of SUSY breaking.
The very existence of the lower bound on $T_R$ implies
that there is a maximal value of a SUSY breaking scale $\mu$ 
which characterizes metastable vacua we can reach as the Universe cools down. 
Using our estimate for $T_R,$ Eq.~\eqref{cond-TR} we find
\begin{equation}
\mu \sim \Lambda_L
\left( \frac{T_R}{\Lambda_L} \right)^{\frac{N_f-N}{2N}},
\end{equation}
where $\Lambda_L$ is the Landau pole of the theory. Thus not only is the 
SUSY breaking determined to lie below $\Lambda_L$, it is also 
parametrically suppressed by powers of $T_R/\Lambda_L$.

\section*{Acknowledgements}   

We thank Sakis Dedes, Stefan Forste and George Georgiou for discussions and valuable comments.
CSC is supported by an EPSRC Advanced Fellowship and VVK by a PPARC Senior Fellowship.

\end{document}